\documentclass[a4paper]{revtex4}
\usepackage{graphicx}
\usepackage{fancyhdr}
\usepackage{amsmath}

\usepackage{color}
\usepackage{graphics}
\usepackage{amsmath, amssymb, epsfig}
\usepackage{subfigure}
\usepackage{graphicx}
\newcommand{\mathsym}[1]{{}}

\newcommand{\ba}{\begin{array}}
\newcommand{\ea}{\end{array}}
\newcommand{\bal}{\begin{align}}
\newcommand{\eal}{\end{align}}
\newcommand{\be}{\begin{equation}}
\newcommand{\ee}{\end{equation}}
\newcommand{\beqa}{\begin{eqnarray}}
\newcommand{\eeqa}{\end{eqnarray}}
\newcommand{ \eq}[1]{Eq.~(\ref{#1})}
\def\321{$SU(3)\times SU(2)\times U(1)$}

\def\ca{\cos\alpha}
\def\cb{\cos\beta}
\def\sa{\sin\alpha}
\def\sb{\sin\beta}

\begin{document}

\title{Muon g--2 in  two-Higgs-doublet models}

\author{Eung Jin Chun}
\email{ejchun@kias.re.kr }
\affiliation{Korea Institute for Advanced Study, Seoul 130-722, Korea}

\begin{abstract}
Updating various theoretical and experimental constraints on the four different types of two-Higgs-doublet models (2HDMs), we find that only the ``lepton-specific" (or  ``type X") 2HDM can explain the present muon g$-$2 anomaly in the parameter region of large $\tan\beta$,
a light CP-odd  boson, and heavier CP-even and charged bosons which are almost degenerate.
The severe constraints on the models come mainly from the consideration of vacuum stability and perturbativity,  the electroweak precision data, the $b$-quark observables like $B_S \to \mu\mu$, the precision measurements of the  lepton universality as well as the 125 GeV  boson property observed at the LHC.
\end{abstract}


\maketitle

\section{Outline}

Since the first measurement of the muon anomalous magnetic moment $a_\mu = (g-2)_\mu/2$
by the E821 experiment at BNL in 2001 \cite{bnl0102}, much progress has been made in both experimental and theoretical sides to reduce the uncertainties by a factor of two or so establishing a consistent $3 \sigma$ discrepancy
\begin{equation} \label{Damu}
\Delta a_\mu \equiv a_\mu^{\rm EXP} - a_\mu^{\rm SM} = + 262\, (85) \times 10^{-11}
\end{equation}
which is in a good agreement with the different group's determinations.
Followed by the 2001 announcement, there have been quite a few studies in the context of 2HDMs  \cite{2hdms,maria0208,cheung0302}, however, restricted mainly to the type I and II models
out of four different types of 2HDMs ensuring natural flavour conservation.
Considering the recent experimental development confirming more precisely
the Standard Model (SM) predictions, including the discovery of the 125 GeV Brout-Egnlert-Higgs boson,
it would be timely to revisit the issue of the muon g--2 in 2HDMs.

An additional contribution to th muon g--2 from an extra boson in 2HDMs, shown in Fig.~1,
may be the origin of the positive excess in the $\Delta a_\mu$.  This can happen in the type II or X (lepton-specific) 2HDM which allows a light boson having large Yukawa couplings enhanced by $\tan\beta$.
While the type II option is completely ruled out by now,  the type X model \cite{cao0909} remains an unique option to explain the $a_\mu$ anomaly evading all the recent experimental constraints \cite{broggio1409,wang1412,abe1504,chun1507}.

\begin{figure}[!ht]
\centering
\subfigure{\includegraphics[width=0.3\textwidth]{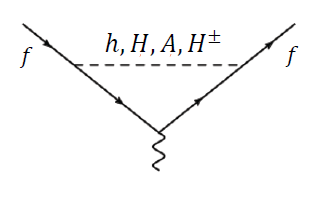}}\quad
\subfigure{\includegraphics[width=0.3\textwidth]{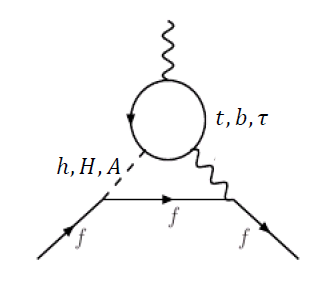}}
\caption{One and two loop diagrams contributing to the muon g$-$2 in 2HDMs.}
\label{fig:loops}
\end{figure}

The previous studies on the muon g--2 in the type II 2HDM (2HDM-II) and various experimental constraints
were nicely summarised in Ref.~\cite{maria0208}:

\begin{itemize}
\item
The one-loop correction mediated by a light CP-even (CP-odd and charged) boson  gives a positive (negative) contribution to $\Delta a_\mu$, and thus the current 3$\sigma$ deviation can be explained by a  CP-even  boson lighter than about 5 GeV.  However, such a light  boson was already  in contradiction to non-observation of radiative $\Upsilon$ decays $\Upsilon\to \gamma +X$.
\item
Contrary to the one-loop correction, the major two-loop contribution from the  Barr-Zee diagram \cite{bz} mediated by a light CP odd (even) boson  can give a sizable positive (negative) contribution to $\Delta a_\mu$.  Thus, a light CP odd boson $A$ with large $\tan\beta$ can account for the muon g--2 deviation.   However,   most of the muon g--2 favoured region in the lower and upper $M_A$ part are excluded  by the LEP and TEVATRON search on $Z \to b \bar{b} A(b\bar{b})$, respectively, except a small gap of $M_A \approx 25-70$ GeV with $\tan\beta \gtrsim 20$.  
\item However, let us note that such a light $A$ gives a huge contribution to $B_s \to \mu^+ \mu^-$ as its rate is proportional to  $\tan^4\beta/M_A^4$ \cite{logan00}, and thus the above gap is completely closed now by the LHC measurement which is consistent with the SM prediction \cite{Bsmumu}.
\end{itemize}

Now the situation  can be quite different in the type X 2HDM (2HDM-X) where
all the extra  boson couplings to quarks (leptons) are  proportional to $\cot\beta$ ($\tan\beta$).
Due to this, the 2HDM-X becomes hadrophobic in the large $\tan\beta$ limit invalidating
most hadron-related constraints applied to the type II model.  On the other hand,
its leptophilic property brings sever constraints from the precision leptonic observables.
The key features in confronting  the type X model with the muon g--2 anomaly can be
summarized as follows \cite{broggio1409,wang1412,abe1504,chun1507}.

\begin{itemize}
\item
As in the 2HDM-II, the one-loop correction with a light CP even boson $H$ can account for the muon g--2 excess. While the Upsilon decay suppressed by $1/\tan^2\beta$ cannot provide a meaningful  constraint, the Belle and LHCb searches for $B\to K \mu \mu$ shut down the muon g--2 favoured region except tiny gaps around $M_H \approx 3$ and $4$ GeV \cite{staub1310}. In any case, such a light Higgs boson is excluded by the current measurement of $B_s \to \mu^+ \mu^-$ even in the 2HDM-X where $\Gamma(B_s \to \mu^+ \mu^-) \propto 1/M_H^4$.
\item
The Barr-Zee diagram with the tau lepton in the loop can account for 
the muon g--2 anomaly again in a parameter region of small $M_A$ and large $\tan\beta$ evading all the constraints from the hardron colliders and the $b$-quark observables \cite{broggio1409} except the process $B_s \to \mu^+ \mu^-$ which rules out  $M_A \lesssim10$ GeV \cite{wang1412}.
\item
However, the lepton universality test by HFAG \cite{hfag} combined with the $Z\to \tau\tau$ decay turns out to limit severely the muon g--2 favoured region of the type X model \cite{abe1504} allowing (only at 2$\sigma$) a small region below $M_A \approx 80$ GeV and $\tan\beta\approx 60$  \cite{chun1507}.
\item
With such a light $A$, the exotic decay of the 125 GeV  boson $h \to AA $ or $A A^*(\tau\tau)$
becomes generically too large unless a certain cancelation is arranged to suppress  the $hAA$ coupling $\lambda_{hAA}$ which turns out to be possible only in the wrong-sign limit of the lepton Yukawa coupling \cite{wang1412}.
\end{itemize}

\section{Four types of 2HDMs}

Non-observation of flavour changing neutral currents restricts 2HDMs to four different classes which differ by how the Higgs doublets couple to fermions~\cite{Gunion:2002zf}. They are organized by a discrete symmetry $Z_2$ under which different Higgs doublets and fermions carry different parities. These models are labeled as type I, II, ``lepton-specific" (or X) and ``flipped" (or Y).  Having two Higgs doublets $\Phi_{1,2}$, the most general $Z_2$ symmetric scalar potential takes the form:
\begin{eqnarray} \label{scalar-potential}
	V &=& m_{11}^2 |\Phi_1|^2
	+ m_{22}^2 |\Phi_2|^2 - m_{12}^2 (\Phi_1^\dagger \Phi_2 + \Phi_1 \Phi_2^\dagger) \nonumber \\
	&& + {\lambda_1\over2} |\Phi_1|^4 + {\lambda_2 \over 2} |\Phi_2|^4
	+ \lambda_3 |\Phi_1|^2 |\Phi_2|^2 + \lambda_4 |\Phi_1^\dagger \Phi_2|^2 + {\lambda_5 \over 2}
	\left[ (\Phi_1^\dagger \Phi_2)^2 + (\Phi_1 \Phi_2^\dagger)^2\right],
\end{eqnarray}
where a (soft) $Z_2$ breaking term $m^2_{12}$ is introduced.
Minimization of the scalar potential determines the vacuum expectation values $\langle \Phi^0_{1,2} \rangle \equiv v_{1,2}/\sqrt{2}$ around which the Higgs doublet fields
are expanded as
\begin{equation}
\Phi_{1,2} = \left[\eta^+_{1,2}, {1\over\sqrt{2}}\left(v_{1,2} + \rho_{1,2} + i \eta^0_{1,2}\right)\right].
\end{equation}
The model contains the five physical fields in mass eigenstates  denoted by $H^\pm, A, H$ and $h$.
Assuming negligible CP violation, $H^\pm$ and $A$ are given by
\begin{equation}
 H^\pm, A = s_\beta\, \eta_1^{\pm, 0}  - c_\beta\, \eta_2^{\pm, 0}
\end{equation}
where the angle $\beta$ is determined from $t_\beta\equiv \tan\beta =v_2/v_1$, and their orthogonal combinations are the corresponding Goldstone modes $G^{\pm, 0}$.
The neutral CP-even bosons are diagonalized as
\begin{equation}
 h = c_\alpha \,\rho_1 - s_\alpha\, \rho_2, \quad
 H = s_\alpha\, \rho_1 + c_\alpha\, \rho_2
\end{equation}
where $h\, (H)$ denotes the lighter (heavier) state.

The gauge couplings of $h$ and $H$ are given schematically by
\begin{equation}
 {\cal L}_{\rm gauge} = g_V m_V \big(s_{\beta-\alpha} h + c_{\beta-\alpha} H \big) VV
\end{equation}
where $V=W^\pm$ or $Z$. Taking $h$ as the 125 GeV boson of the SM, the SM limit corresponds to
$s_{\beta-\alpha} \to 1$.  Indeed, LHC finds,  $c_{\beta-\alpha} \ll 1 $ in all the 2HDMs confirming the SM-like property of the 125 GeV boson \cite{atlas15}.

\begin{table}[!ht]
\begin{small}
\begin{center}
\begin{tabular}{|l|ccccccccc|}
 \hline
~~~~~~~~&~~~$y_u^A$~~~ & ~~~$y_d^A$~~~ & ~~~$y_l^A$~~~ & ~~~~$y_u^H$~~~ & ~~~$y_d^H$~~~ & ~~~$y_l^H$~~~ & ~~~$y_u^h$~~~ & ~~~$y_d^h$~~~ & ~~~$y_l^h$~~~\\
\hline
~Type I~~ &$\cot\beta$ & $-\cot\beta$ & $-\cot\beta$ & $\frac{\sa}{\sb}$ & $\frac{\sa}{\sb}$ & $\frac{\sa}{\sb}$ & $\frac{\ca}{\sb}$ & $\frac{\ca}{\sb}$ & $\frac{\ca}{\sb}$~ \\
~Type II~~ &$\cot\beta$ & $\tan\beta$ & $\tan\beta$ & $\frac{\sa}{\sb}$ & $\frac{\ca}{\cb}$ & $\frac{\ca}{\cb}$ & $\frac{\ca}{\sb}$ & $-\frac{\sa}{\cb}$ & $-\frac{\sa}{\cb}$~ \\
~Type X~~ &$\cot\beta$ & $-\cot\beta$ & $\tan\beta$ & $\frac{\sa}{\sb}$ & $\frac{\sa}{\sb}$ & $\frac{\ca}{\cb}$ & $\frac{\ca}{\sb}$ & $\frac{\ca}{\sb}$ & $-\frac{\sa}{\cb}$~ \\
~Type Y~~ &$\cot\beta$ & $\tan\beta$ & $-\cot\beta$ & $\frac{\sa}{\sb}$ & $\frac{\ca}{\cb}$ & $\frac{\sa}{\sb}$ & $\frac{\ca}{\sb}$ & $-\frac{\sa}{\cb}$ & $\frac{\ca}{\sb}$~ \\
\hline
\end{tabular}
\end{center}
\end{small}
\caption{The normalized Yukawa couplings for up- and down-type quarks and charged leptons.}
\label{yukawas}
\end{table}

Normalizing the Yukawa couplings of the neutral bosons to a fermion $f$  by $m_f/v$ where $v=\sqrt{v_1^2+v_2^2} = 246$ GeV, we have the following Yukawa terms:
\begin{eqnarray}
-{\cal L}^{\rm 2HDMs}_{\rm Yukawa} &=&
\sum_{f=u,d,l} {m_f\over v}
\left( y_f^h h \bar{f} f + y_f^H H \bar{f} f - i y_f^A A \bar{f} \gamma_5 f \right) \\
&& \nonumber
+\left[ \sqrt{2} V_{ud} H^+  \bar{u} \left( {m_u\over v} y^A_u P_L  + {m_d \over v} y^A_d P_R\right) d
+\sqrt{2} {m_l \over v} y_l^A H^+ \bar{\nu} P_R l + h.c.\right]
\end{eqnarray}
where the normalized Yukawa couplings $y^{h,H,A}_f$ are summarized in Table I for each of these four types of 2HDMs.

Let us now recall that the tau Yukawa coupling $y_\tau \equiv y^h_l$ in Type X
(also $y_b \equiv y^h_d$ in Type II)
can be expressed as
\begin{equation} \label{ytau}
y_\tau = -{s_\alpha \over c_\beta} = s_{\beta-\alpha} - t_\beta c_{\beta-\alpha}
\end{equation}
which allows us to have the wrong-sign limit $y_\tau \sim -1$ compatible with the LHC data \cite{ferreira1410}  if $c_{\beta-\alpha}\sim 2/t_\beta$ for large $t_\beta\equiv \tan\beta$ favoured by the muon g$-$2.
Later we will see that a cancellation in $\lambda_{hAA}$ can be arranged only for $y^h_\tau<-1$
to suppress the $h\to AA$ decay.

\section{The Muon $\mbox{\bf g}$$-$2 from a light CP-odd boson}
\label{sec:gm2}

\begin{figure}[!ht]
\centering
\subfigure{\includegraphics[width=0.35\textwidth]{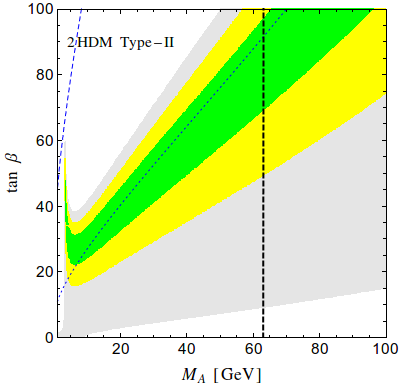}}\quad
\subfigure{\includegraphics[width=0.35\textwidth]{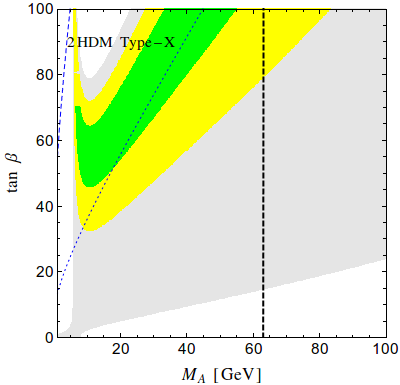}}
\caption{The $1\sigma$, $2\sigma$ and $3\sigma$ regions allowed by $\Delta a_\mu$ in the
$M_A$-$\tan\beta$ plane taking the limit of $c_{\beta-\alpha}=0$ and $M_{h(H)}=125$ (200) GeV in
type II (left panel) and type X (right panel) 2HDMs. The regions below the dashed (dotted) lines are allowed at 3$\sigma$ (1.4$\sigma$) by $\Delta a_e$. The vertical
dashed line corresponds to $M_A = M_h/2$. }
\label{fig:g-2}
\end{figure}

Considering all the updated SM calculations of the muon g$-$2,  we obtain
\be
  a_{\mu}^{\mbox{$\scriptscriptstyle{\rm SM}$}}= 116591829 \, (57) \times 10^{-11}
\ee
comparing it with the experimental value
$   a_{\mu}^{\mbox{$\scriptscriptstyle{\rm EXP}$}}  = 116592091 \, (63) \times 10^{-11}$,
one finds a deviation at 3.1$\sigma$:
$\Delta a_{\mu} \equiv a_{\mu}^{\mbox{$\scriptscriptstyle{\rm EXP}$}}-a_{\mu}^{\mbox{$\scriptscriptstyle{\rm SM}$}} = +262 \, (85) \times 10^{-11}$.
In the 2HDM, the one-loop contributions to $a_{\mu}$ of the neutral and charged 
bosons are
\be
	\Delta a_\mu^{\mbox{$\scriptscriptstyle{\rm 2HDM}$}}({\rm 1loop}) =
	\frac{G_F \, m_{\mu}^2}{4 \pi^2 \sqrt{2}} \, \sum_j  \left (y_{\mu}^j \right)^2  r_{\mu}^j \, f_j(r_{\mu}^j),
\label{amuoneloop}
\end{equation}
where $j =  \{h, H, A , H^\pm\}$, $r_{\mu}^ j =  m_\mu^2/M_j^2$, and
\begin{eqnarray}
	f_{h,H}(r) &=& \int_0^1 \! dx \,  { x^2 ( 2- x) \over 1 - x + r x^2},
\label{oneloopintegrals1} \\
	f_A (r) &=& \int_0^1 \! dx \,  { -x^3  \over 1 - x +r  x^2},
\label{oneloopintegrals2} \\
	f_{H^\pm} (r) &=& \int_0^1 \! dx \, {-x (1-x)  \over 1 - (1-x) r}.
\label{oneloopintegrals3}
\end{eqnarray}
These formula show that the one-loop contributions to $a_{\mu}$ are positive for the neutral scalars $h$ and $H$, and negative for the pseudo-scalar and charged  bosons $A$ and $H^{\pm}$ (for $M_{H^\pm} > m_{\mu}$). In the limit $r\ll1$,
\begin{eqnarray}
	f_{h,H}(r) &=&- \ln r - 7/6 + O(r),
	\label{oneloopintegralsapprox1} \\
	f_A (r) &=& +\ln r +11/6 + O(r),
	\label{oneloopintegralsapprox2} \\
	f_{H^\pm} (r) &=& -1/6 + O(r),
	\label{oneloopintegralsapprox3}
\end{eqnarray}
showing that in this limit $f_{H^\pm}(r)$ is suppressed with respect to $f_{{h,H,A}}(r)$.
Now the two-loop Barr-Zee type diagrams with effective
$h\gamma \gamma$, $H\gamma \gamma$ or  $A\gamma \gamma$ vertices generated
by the exchange of heavy fermions gives
\be
	\Delta a_\mu^{\mbox{$\scriptscriptstyle{\rm 2HDM}$}}({\rm 2loop-BZ}) = \frac{G_F \, m_{\mu}^2}{4 \pi^2 \sqrt{2}} \, \frac{\alpha_{\rm em}}{\pi}
	\, \sum_{i,f}  N^c_f  \, Q_f^2  \,  y_{\mu}^i  \, y_{f}^i \,  r_{f}^i \,  g_i(r_{f}^i),
\label{barr-zee}
\end{equation}
where $i = \{h, H, A\}$, $r_{f}^i =  m_f^2/M_i^2$, and $m_f$, $Q_f$ and $N^c_f$ are the mass, electric charge and number of color degrees of freedom of the fermion $f$ in the loop. The functions $g_i(r)$ are
\be
\label{2loop-integrals}
	g_i(r) = \int_0^1 \! dx \, \frac{{\cal N}_i(x)}{x(1-x)-r} \ln \frac{x(1-x)}{r},
\ee
where ${\cal N}_{h,H}(x)= 2x (1-x)-1$ and ${\cal N}_{A}(x)=1$.

Note the enhancement factor $m_f^2/m_{\mu}^2$ of the two-loop formula in~\eq{barr-zee} relative to the one-loop contribution in~\eq{amuoneloop},
which can overcome the additional loop suppression factor $\alpha / \pi$, and makes
the two-loop contributions may  become larger than the one-loop ones.
Moreover, the signs of the two-loop functions $g_{h,H}$ (negative) and
$g_{A}$ (positive) for the CP-even and CP-odd contributions are
opposite to those of the functions $f_{h,H}$ (positive) and $f_{A}$ (negative) at one-loop.
As a result, for small $M_A$ and large $\tan \beta$ in Type II and X, the positive two-loop pseudoscalar contribution can generate a dominant contribution which can account for
the observed $\Delta a_{\mu}$ discrepancy.
The additional 2HDM contribution $\delta a_{\mu}^{\mbox{$\scriptscriptstyle{\rm 2HDM}$}}  = \delta a_\mu^{\mbox{$\scriptscriptstyle{\rm 2HDM}$}}({\rm 1loop}) + \delta a_\mu^{\mbox{$\scriptscriptstyle{\rm 2HDM}$}}({\rm 2loop-BZ})$ obtained adding Eqs.~(\ref{amuoneloop}) and~(\ref{barr-zee}) (without the $h$ contributions) is compared with $\Delta a_\mu$ in Fig.~\ref{fig:g-2}.

\section{Electroweak Precesion Data}
\label{sec:ewc}

\begin{figure}[!ht]
\centering
\includegraphics[width=1.0\textwidth]{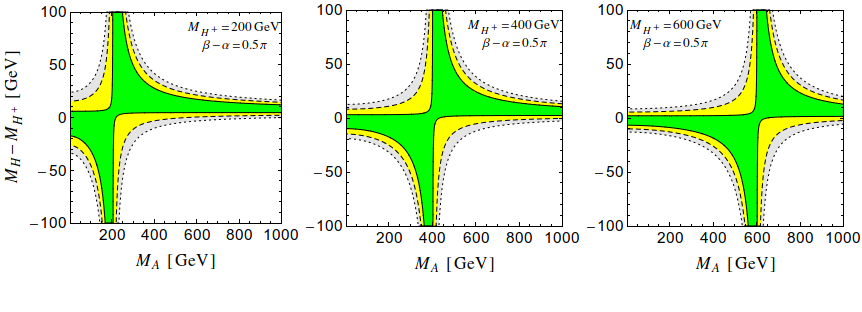}
\caption{The parameter space allowed in the $M_A$ vs.\ $\Delta M_H =M_H-M_{H^\pm}$ plane by the electroweak precision constraints. The green, yellow, gray regions satisfy $\Delta \chi_{\mbox{$\scriptscriptstyle{\rm EW}$}}^2 (M_A, \Delta M)< 2.3, 6.2, 11.8$, corresponding to 68.3, 95.4, and 99.7\% confidence intervals, respectively.}
\label{fig:ewpc}
\end{figure}

Allowing such a light CP-odd boson,
there could be a strong limit on the extra boson masses
coming from the electroweak precision test.
To see this, we compare the theoretical 2HDMs predictions for $M_W$ and $\sin^2\!\theta^{\text{lept}}_{\text{eff}}$ with their present experimental values via a combined $\chi^2$ analysis. These quantities can be computed perturbatively by means of the following relations
\begin{eqnarray}
	M^2_W &=& \frac{M^2_Z}{2}\left[1 + \sqrt{1 - \frac{4 \pi \alpha_{\rm em}}{\sqrt{2} G_F M^2_Z} \frac{1}{1 - \Delta r}} \, \right]\\
	\sin^2\!\theta^{\text{lept}}_{\text{eff}} &=&
 k_l \! \left(M^2_Z\right) \sin^2\!\theta_W \, ,
\end{eqnarray}
where $\sin^2\!\theta_W=1-M^2_W/M^2_Z$, and $k_l(q^2) = 1+\Delta k_l(q^2)$ is the real part of the vertex form factor $Z\to l \bar{l}$ evaluated at $q^2 = M^2_Z$.
We than use the following experimental values:
\beqa \label{exp}
	M_W^{\mbox{$\scriptscriptstyle{\rm EXP}$}} &=& 80.385 \pm 0.015~{\rm GeV}, \nonumber \\
	\sin^2\!\theta^{\text{lept}, \scriptscriptstyle{\rm EXP}}_{\text{eff}} &=& 0.23153 \pm 0.00016.
	\eeqa
The results of our analysis are displayed in Fig.~\ref{fig:ewpc} confirming a custodial symmetry
limit  \cite{gerard0703} of our interest $M_A \ll M_H \sim M_{H^\pm}$, $M_A\sim M_{H^\pm} \ll M_H$, or $M_A \sim M_H \sim M_{H^\pm}$ although the last two cases are disfavoured.

\section{Theoretical Consideration of vaccum stability and perturbativity}
\label{sec:thc}

While any value of $M_A$ is allowed by the EW precision tests in the limit of $M_H \sim M_{H^\pm}$, there appear upper bounds on $M_{H^\pm, H}$ by theoretical consideration of vacuum stability, global minimum, and perturbativity expressed respectively by
\begin{eqnarray} \label{vacuum-stability}
&&	\lambda_{1,2}>0,~~\lambda_3>-\sqrt{\lambda_1
	\lambda_2},~~|\lambda_5|<\lambda_3+\lambda_4+\sqrt{\lambda_1 \lambda_2}, \\
&& m_{12}^2(m_{11}^2-m_{22}^2\sqrt{\lambda_1/\lambda_2})(\tan\beta-(\lambda_1/\lambda_2)^{1/4})>0, \\
&&	|\lambda_i| \lesssim |\lambda_{\rm max}| =  \sqrt{4\pi}, 2\pi,\mbox{ or } 4\pi .
\end{eqnarray}
Taking $\lambda_1$ as a free parameter, one can have the following expressions for the other couplings in the large $t_\beta$ limit:
\begin{eqnarray} \label{lambdas}
\lambda_2 v^2 &\approx& s^2_{\beta-\alpha} M_h^2  \\
\lambda_3 v^2 &\approx& 2 M^2_{H^\pm}
-(s^2_{\beta-\alpha} + s_{\beta-\alpha} y_\tau) M_H^2 + s_{\beta-\alpha} y_\tau M_h^2\\
\lambda_4 v^2 &\approx&  -2 M^2_{H^\pm} + s^2_{\beta-\alpha} M^2_H + M_A^2 \\
\lambda_5 v^2 &\approx&  s^2_{\beta-\alpha} M^2_H - M_A^2
\end{eqnarray}
where we have used the relation (\ref{ytau}) neglecting the terms of
${\cal O}(1/t_\beta^2)$.

In the right-sign (RS) limit of the lepton (tau, in particular), $y_\tau s_{\beta-\alpha} \to +1$, one finds a strong upper limit of  \cite{broggio1409}
\begin{equation}
 M_A \ll M_{H^\pm} \sim M_H \lesssim 250 \mbox{GeV} \quad \mbox{(RS)}.
\end{equation}
On the other hand,  in the wrong-sing (WS) limit, $y_\tau s_{\beta-\alpha} \to -1$,
  the heavy  boson masses up to the perturbativity limit,
\begin{equation}
 M_A \ll M_{H^\pm} \sim M_H \lesssim \sqrt{4\pi} v \quad \mbox{(WS)}.
\end{equation}
can be obtained.

 \medskip

Let us finally remark that the $hAA$ coupling is generically order one and thus can leads to a sizable non-standard decay of $h \to A A$ or $AA^*(\tau\tau)$ if allowed kinematically. Then, one needs to  have $|\lambda_{hAA}/v|\ll1$ to avoid an exotic decay of the SM  boson.
Noting that  $\lambda_{hAA}/v \approx s_{\beta-\alpha} [\lambda_3+\lambda_4-\lambda_5]$, one gets
\begin{equation}
\lambda_{hAA} v/s_{\beta-\alpha} \approx
-(1+s_{\beta-\alpha} y_\tau) M_H^2 + s_{\beta-\alpha} y_\tau M_h^2 + 2 M_A^2
\end{equation}
where we have put $s^2_{\beta-\alpha}=1$. In the RS or SM limit, the condition $\lambda_{hAA} \approx 0$ can be met for a rather light $H$ with $M_H^2 \approx  {1\over2} M_h^2 +M_A^2$ which is disfavoured in the explanation of the muon g--2.  On the other hand, one can arrange a cancellation for $\lambda_{hAA} \approx 0$ in the wrong-sign limit  for arbitrary value of $M_H$ if the tau Yukawa coupling satisfies
\begin{equation}
y_\tau s_{\beta-\alpha} \approx - {M_H^2 - 2 M_A^2 \over M_H^2-M_h^2} < -1.
\end{equation}

\section{Lepton universality tests}

In the limit of large $\tan\beta$, the charged boson can generate
significant corrections to $\tau$ decays at the tree level and furthermore the extra Higg boson contribution to one-loop corrections can also be significant~\cite{maria04}.
The recent study \cite{abe1504} showed that a stringent bound on the charged boson contributions can be obtained from the lepton universality condition obtained by the HFAG collaboration \cite{hfag}.
Given the precision at the level of  0.1 \%, the lepton universality data put the strongest bound on the type X 2HDM parameter space in favor of the muon g--2.  Thus, let us now make a proper analysis of the HFAG data.

From the measurements of the pure leptonic processes, $\tau \to \mu \nu \nu$, $\tau \to e \nu \nu$
 and $\mu \to e \nu\nu$, HFAG obtained the constraints on the three coupling ratios,
 $(g_\tau/g_\mu) = \sqrt{ \Gamma( \tau \to e  \nu\nu)/\Gamma(\mu \to e \nu\nu)}$, etc.
Defining $\delta_{l l'}\equiv (g_l/g_{l'})-1$, let us rewrite the data from the leptonic processes:
\begin{equation} \label{LUdata1}
\delta^l_{\tau\mu} =  0.0011 \pm 0.0015,\quad
\delta^l_{\tau e} = 0.0029 \pm 0.0015, \quad
\delta^l_{\mu e} = 0.0018 \pm 0.0014
\end{equation}
In addition, combing the semi-hadronic processes $\pi/K \to \mu \nu$, HFAG  also provided
the averaged constraint on $(g_\tau/g_\mu)$ which is translated into
\begin{equation} \label{LUdata2}
\delta^{l+\pi+K}_{\tau\mu} = 0.0001 \pm 0.0014.
\end{equation}

It is  important to notice that only two ratios out of the three leptonic measurements are independent
and thus the three data (\ref{LUdata1}) are strongly correlated. For a consistent treatment of the data, one combination out of the three  has to be projected out. One can  indeed check that the direction $\delta^l_{\tau\mu}-\delta^l_{\tau e} + \delta^l_{\mu e}$ has the zero best-fit value and the zero eigenvalue of the covariance matrix, and thus corresponds to the unphysical direction.
Furthermore, two orthogonal directions $\delta^l_{\tau\mu} + \delta^l_{\tau e}$ and $-\delta^l_{\tau\mu}+\delta^l_{\tau e} + 2 \delta^l_{\mu e}$ are found to be uncorrelated in a good approximation.  As a result, the 2HDM contribution to  $\delta_{l l'}$ are calculated to be
\begin{equation} \label{deltas}
\delta^l_{\tau\mu} = \delta_{loop},\quad
\delta^l_{\tau e} = \delta_{tree} + \delta_{loop},\quad
\delta^l_{\mu e} = \delta_{tree}, \quad
\delta^{l+\pi+K}_{\tau \mu} = \delta_{loop}.
\end{equation}
Here $\delta_{tree}$ and $\delta_{loop}$ are given by \cite{maria04}:
\begin{eqnarray} \label{delta_tree_loop}
\delta_{tree} &=& {m_\tau^2 m_\mu^2 \over 8 m^4_{H^\pm}} \tan^4 \beta
- {m_\mu^2 \over m^2_{H^\pm}} t^2_\beta {g(m_\mu^2/m^2_\tau) \over f(m_\mu^2/m_\tau^2)}, \\
\delta_{loop} &=& {G_F m_\tau^2 \over 8 \sqrt{2} \pi^2} t^2_\beta
\left[1 + {1\over4} \left( H(x_A) + s^2_{\beta-\alpha} H(x_H) + c^2_{\beta-\alpha} H(x_h)\right)
\right]. \nonumber
\end{eqnarray}
where $f(x)\equiv 1-8x+8x^3-x^4-12x^2 \ln(x)$, $g(x)\equiv 1+9x-9x^2-x^3+6x(1+x)\ln(x)$,
$H(x) \equiv \ln(x) (1+x)/(1-x)$, and $x_\phi=m_\phi^2/m_{H^{\pm}}^2$.
From Eqs.~(\ref{LUdata1}), (\ref{LUdata2}) and (\ref{deltas}), one obtains the following three independent bounds:
\begin{eqnarray} \label{LUconstraints}
 {1\over\sqrt{2}} \delta_{tree} +\sqrt{2} \delta_{loop} &=& 0.0028 \pm 0.0019,\quad \nonumber\\
{\sqrt{3\over2} } \delta_{tree} &=& 0.0022\pm 0.0017,\quad \\
\delta_{loop} &=&0.0001 \pm 0.0014 \nonumber.
\end{eqnarray}
We will use these constraints to put a strong limit on the $(g-2)_\mu$ favoured region in the $M_A$--$\tan\beta$ plane in the next section.  Let us recall that the $Z \to \tau \tau$ data, although less strong than the HFAG data, provides an independent bound \cite{abe1504} which further cuts out some corner of parameter space.

\section{Pinning down the whole 2HDM-X parameter space}

It is an interesting task to narrow down the allowed region of  
the type X 2HDM parameter space collecting all the relevant experimental data including those outlined in Section I and the 125 GeV boson data 
from LHC, in particular. The scan ranges of all the 2HDM-X input parameters are listed in Table \ref{tab:inputs}. For our scan, we adopt the convention $-\pi/2<\alpha-\beta<\pi/2$ and $0<\beta<\pi/2$, and use
the parameter $\lambda_1$ as an input parameter instead of $m^2_{12}$.

\begin{table}[h!]
\begin{center}
\begin{tabular}{|c|c|}
\hline
2HDM parameter & Range \\
\hline
Scalar boson mass (GeV) &  $125<m_H<{400}$ \\
Pseudoscalar boson mass (GeV) & $10<m_A<400$ \\
Charged boson mass (GeV) & $94<m_{H^\pm}<400$ \\
$ c_{\beta-\alpha}$& {$0.0< c_{\beta-\alpha}<0.1$} \\
$ \tan\beta$ & $10< \tan\beta <150$ \\
$\lambda_1$ & $0.0<\lambda_1<4\pi$ \\
\hline
\end{tabular}
\caption{The scan ranges of the 2HDM-X input parameters.
}
\label{tab:inputs}
\end{center}
\end{table}

\begin{figure}[!ht]
\centering
\subfigure{\includegraphics[width=0.4\textwidth]{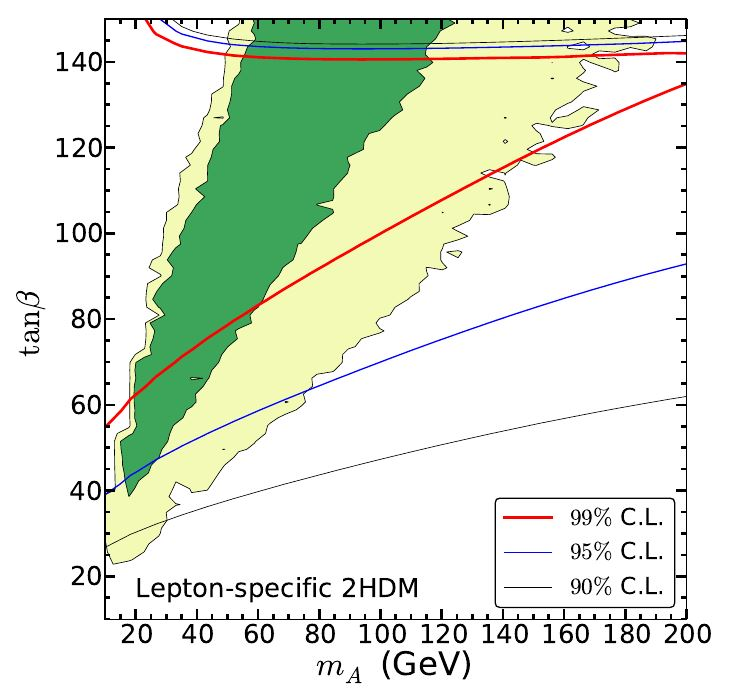}}
\caption{The 68\% and 95\% allowed regions by $\Delta a_\mu$ and other experimental constraints in the $m_A$-$\tan\beta$ plane. The contours of the lepton universality likelihood in the 99\%, 95\% and 90\% confidence level are overlayed.   }
\label{fig:g-2}
\end{figure}

Fig.~\ref{fig:g-2} shows the allowed region in the $m_A$-$\tan\beta$ plane from the profile-likelihood study  taking all the other 2HDM-X parameters as nuisance parameters. To see the impact of the lepton universality data by HFAG, we overlay the contour lines of the lepton universality likelihood at the 99\%, 95\% and 90\% confidence level based on the constraints (\ref{LUconstraints}).  The allowed region opened up for $\tan\beta>140$ needs a comment.
Note that the $\delta_{loop}$ is always negative while $\delta_{tree}$  becomes positive for larger  $\tan\beta/m_{H^\pm}$. Thus, there appears a fine-tuned region around $\tan\beta/m_{H^\pm} \sim 1$~GeV$^{-1}$ where the positive $\delta_{tree}$ and the negative $\delta_{loop}$ cancel each other to give a good fit. However, such regions are excluded by the $Z\to \tau\tau$ data \cite{abe1504} and thus we are left with the tightly limited region of $M_A \approx 10-80$ GeV and $\tan\beta\approx 25-60$ at the 95\% confidence level.

\begin{figure}[!ht]
\centering
\subfigure{\includegraphics[width=0.3\textwidth]{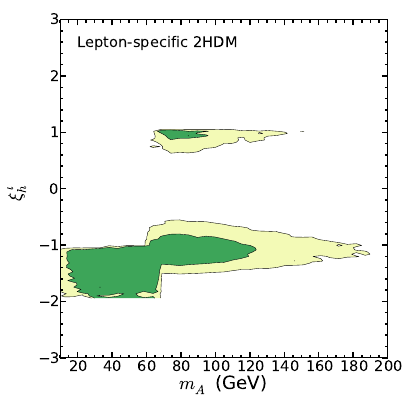}}\quad
\subfigure{\includegraphics[width=0.3\textwidth]{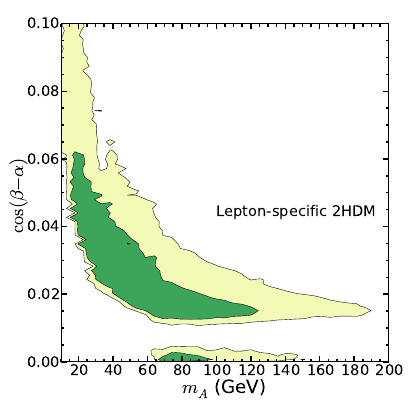}}\quad
\subfigure{\includegraphics[width=0.3\textwidth]{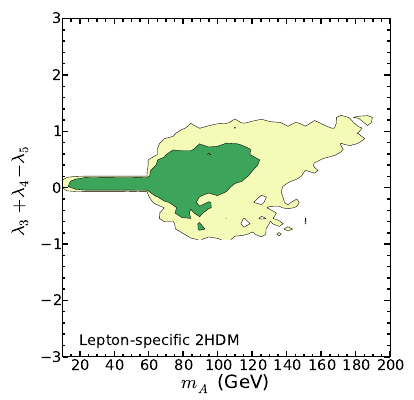}}\quad
\caption{The 68\% and 95\% allowed regions in the $m_A$--$\xi^l_h$  ($y_\tau\equiv \xi^l_h$) (left) and
$m_A$--$c_{\beta-\alpha}$ (middle), and $m_A$--$\lambda_{hAA}$ ($\lambda_{hAA} \approx \lambda_3+\lambda_4-\lambda_5$) (right) plane.}
\label{fig:RW}
\end{figure}

The region allowed in Fig.~\ref{fig:g-2} can be either in the right-sign  ($y_\tau \equiv \xi^l_h>0$) or
wrong-sign  ($y_\tau \equiv \xi^l_h <0 $) domain as shown in the left and middle panels of  Fig.~\ref{fig:RW}. One can see that the rigt-sign limit is tightly constrained to a small region of   $m_A\approx 60-80 $ GeV while the wrong-sign limit is favoured in a wider range of parameter space.
The right panel shows the sizes of the coupling $\lambda_{hAA}$ restricted by the LHC data on the exotic decay of the 125 GeV  boson, putting a generous bound of  Br($h\to AA^{(*)}$) $<$ 40\%. As explained before, the suppressed value of $\lambda_{hAA}$ for $m_A \lesssim m_h/2$ is shown to appear only in the wrong-sign domain.

\begin{figure}[!ht]
\centering
\subfigure{\includegraphics[width=0.35\textwidth]{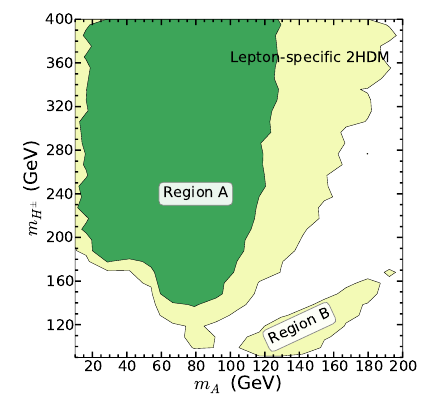}}\quad
\subfigure{\includegraphics[width=0.34\textwidth]{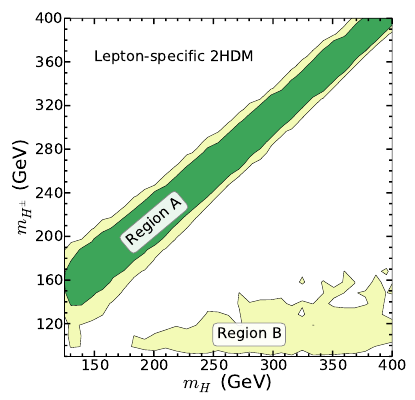}}
\caption{Distribution of the extra boson masses allowed at the 68\% and 95\% confidence levels. }
\label{fig:AHHc}
\end{figure}

Given the possible existence of a light CP-odd  boson explaining the muon g--2 in the type X 2HDM, it would be important to look for its trails at the LHC.
Fig.~\ref{fig:AHHc} shows the allowed mass ranges of all the extra  bosons. Region A following the pattern of $m_A \lesssim m_H \approx m_{H^\pm}$  is favoured while Region B with  $m_A \approx m_{H^\pm} \ll m_H$ is already excluded as discussed before.

\section{Tau-rich sigantures at the LHC}

The bulk parameter space with $m_A \ll m_H \sim m_{H^\pm}$   is a clear prediction of the type X 2HDM as the origin of the muon g--2 anomaly.  Since the extra  bosons are mainly from
the ``leptonic'' Higgs doublet with  large $ \tan \beta$, all the three members are expected to dominantly decay into the $\tau-$flavor, leading to $\tau-$rich signatures at the LHC via the following production and ensuing cascade decay chains:
\begin{align}
pp\to& W^{\pm *} \to H^\pm A \to (\tau^\pm \nu)(\tau^+\tau^-),\\
pp\to  &       Z^*/\gamma^* \to HA \to  (\tau^+\tau^-)(\tau^+\tau^-), \nonumber\\
pp\to& W^{\pm *} \to H^\pm H \to (\tau^\pm \nu)(\tau^+\tau^-),\nonumber \\
pp\to& Z^*/\gamma^* \to H^+H^- \to (\tau^+\nu)(\tau^-\bar\nu). \nonumber
\end{align}
 To probe Region A, we select six benchmark points with different combinations of $m_A$ and $m_H$ presented in Table~\ref{tab:selectioncut2}.
For each point, we take a simple parametrization of $\tan\beta = 1.25 (m_A/\mbox{GeV})+25$ and
$m_{H^\pm} = m_H + 15 \mbox{GeV}$.  Note that we included the points with $m_A >80$ GeV  for the sake of the LHC study although they are forbidden by the lepton universality tests.  In Table~\ref{tab:selectioncut2}, we show the production cross-section, the selection cuts and the significance for each benchmark expected for the integrated luminosity of 25/fb at the 14 TeV LHC.

\begin{table}[htb]
\begin{tabular}{l|rrrrrr}
 & point A & point B & point C & point D & point E & point F \cr
\hline
$m_A$~[GeV] & 20 & 40 & 100 & 40 & 100 & 180  \cr
$m_H$~[GeV] & 200 & 200 & 200 & 260 & 260 & 260  \cr
\hline
\hline
total $\sigma_{\rm gen}$ [fb] & 270.980 & 241.830 & 153.580 & 100.430 & 71.271 & 44.163\cr
\hline
$n_{\ell} \ge 3$  & 6.606 & 16.681 & 21.713 & 7.110 & 11.962 & 8.822\cr
$n_{\tau} \ge 3$  & 0.894 & 2.602& 4.386 & 0.888 & 2.346 & 1.971\cr
$E\!\!\!/_T > 100$~GeV    & 0.201 & 0.547 & 1.179 & 0.209 & 0.765 & 0.926\cr
$n_b=n_{j}=0$& 0.098 & 0.314  & 0.857 & 0.121  & 0.479 & 0.631\cr
\hline
\hline
$S/B$ & 0.1  & 0.5 & 1.2 & 0.2 &0.7 & 0.9 \cr
$S/\sqrt{B}_{25{\rm fb}^{-1}}$ & 0.6 & 1.9 & 5.2 & 0.7 & 2.9 & 3.8\cr
\hline
\end{tabular}
\caption{The number of events after applying successive cuts for 14 TeV LHC. }
\label{tab:selectioncut2}
\end{table}

\begin{figure}[!ht]
\centering
\subfigure{\includegraphics[width=0.35\textwidth]{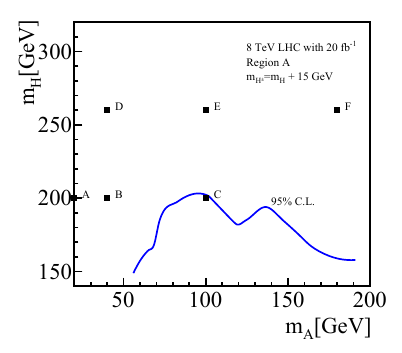}}\quad
\subfigure{\includegraphics[width=0.34\textwidth]{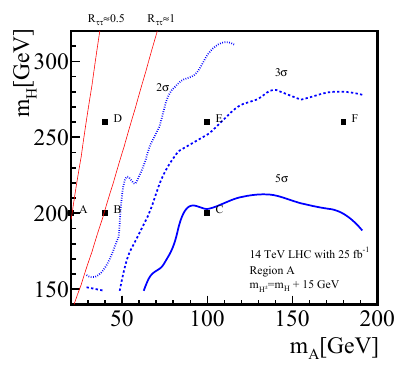}}
\caption{The 95\% exclusion contour from LHC8  (left) and the 2$\sigma$, 3$\sigma$ and 5$\sigma$ discovery reach countours at LHC14 (right) in the $m_A$--$m_H$ plane. }
\label{fig:lhc-regionA}
\end{figure}

In Fig.~\ref{fig:lhc-regionA}, we present the exclusion region coming mainly from the chargino-neutralino search at the  LHC8, and the expected discovery reaches at LHC14 with the integrated luminosity of 25/fb.
A heavy  CP-even  boson with $m_H > 200$~GeV  and a light CP-odd  boson with $m_A<50$~GeV are still allowed, and the LHC14 can explore
some of the regions. The sensitivities are weaker for larger $m_H$ just because of smaller cross sections,
and for smaller $m_A$ because $\tau$s from lighter $A$ become softer and thus the acceptance quickly decreases. Moreover, the $H/H^\pm \to A Z/W^\pm$ decay modes also
start open to decrease the number of hard $\tau$s from direct $H/H^\pm$ decays.
In such a region,  a light $A$ from heavy $H^+/H$ decay will be boosted,
resulting in a collimated $\tau-$pair which becomes difficult to be tagged as two separated $\tau$-jets.
It is one of the reasons to have less acceptance for this parameter region.
We can estimate the separation $R_{\tau\tau}$ of the $\tau$ leptons from $A$ decay:
\begin{equation}
R_{\tau\tau} \sim \frac{2m}{p_T} \sim \frac{4m_A}{m_{H^\pm/H}\sqrt{1 - 2\frac{m_A^2 + m_{W/Z}^2}{m_{H^\pm/H}^2} + \frac{(m_A^2 - m_{W/Z}^2)^2}{m_{H^\pm/H}^4} }}.
\end{equation}
Since the jets are usually defined with $R=0.5$, the $\tau-$pair starts overlapping.
We indicated the region with the overlapping $\tau$ problem in red lines in the right panel of Fig.~\ref{fig:lhc-regionA}. Further studies on
how to capture the kinematic features of the boosted $A\to\tau^+\tau^-$ are required to probe such a
small $m_A$ region.

\section{Summary}
The type X 2HDM is still a viable option for the explanation of the muon  g--2 in the parameter region with large $\tan\beta$ and a light CP-odd  boson $A$. Being ``hadrophobic and leptophilic''  in the large $\tan\beta$ limit, it can be easily free from all the hadron-related constraints, particularly, coming from the decay $B_s\to \mu\mu$ which puts only  a mild bound of $m_A\gtrsim 10$ GeV.  
However, such a region is tightly limited by the lepton universality tests from the HFAG and $Z\to \tau\tau$ data. Combining all the current bounds, we find allowed at the 95\% confidence level a limited region of $\tan\beta\approx 15-60$ and $m_A \approx10-80$ GeV with $m_H \approx m_{H^\pm}\gg m_A$.

It will be an interesting task to search for such a light CP-odd boson $A$ and the extra heavy bosons $H, H^\pm$ in the next run of the LHC mainly through $pp \to  A H, AH^\pm$ followed by the decays $H^\pm \to \tau^\pm \nu$ and $A,H\to \tau^+ \tau^-$ which requires further studies to
improve the (boosted) tau identification.


\end{document}